\documentclass[10pt,aps,prl,twocolumn,preprintnumbers,amsmath,amssymb,floatfix,showpacs,citeautoscript,superscriptaddress]{revtex4-2} 
\usepackage{hyperref}
\usepackage{graphicx}
\usepackage{amsmath}

\usepackage{booktabs} 

\usepackage{color}
\usepackage{siunitx}
\usepackage{makecell}

\begin{document}

\title{Guided Diffusion for the Discovery of New Superconductors}

\author{Pawan Prakash}
\affiliation{Department of Physics, University of Florida}
\affiliation{Quantum Theory Project, University of Florida}
\author{Jason B. Gibson}
\affiliation{Quantum Theory Project, University of Florida}
\affiliation{Department of Materials Science and Engineering, University of Florida}
\author{Zhongwei Li}
\affiliation{Department of Physics, University of Florida}
\author{Gabriele Di Gianluca}
\affiliation{Department of Physics, University of Florida}
\author{Juan Esquivel}
\affiliation{Department of Physics, University of Florida}
\author{Eric Fuemmeler}
\affiliation{Department of Aerospace Engineering and Mechanics, University of Minnesota}
\author{Benjamin Geisler}
\affiliation{Department of Physics, University of Florida}
\affiliation{Department of Materials Science and Engineering, University of Florida}
\author{Jung Soo Kim}
\affiliation{Department of Physics, University of Florida}
\author{Adrian Roitberg}
\affiliation{Quantum Theory Project, University of Florida}
\affiliation{Department of Chemistry, University of Florida}
\author{Ellad B. Tadmor}
\affiliation{Department of Aerospace Engineering and Mechanics, University of Minnesota}
\author{Mingjie Liu}
\affiliation{Quantum Theory Project, University of Florida}
\affiliation{Department of Chemistry, University of Florida}
\author{Stefano Martiniani}
\affiliation{Center for Soft Matter Research, Department of Physics, New York University}
\affiliation{Simons Center for Computational Physical Chemistry, New York University}
\affiliation{Courant Institute of Mathematical Sciences, New York University}
\affiliation{Center for Neural Science, New York University}
\author{Gregory R. Stewart}
\affiliation{Department of Physics, University of Florida}
\author{James J. Hamlin}
\affiliation{Department of Physics, University of Florida}
\author{Peter J. Hirschfeld}
\affiliation{Department of Physics, University of Florida}
\author{Richard G. Hennig}
\affiliation{Department of Physics, University of Florida}
\affiliation{Quantum Theory Project, University of Florida}
\affiliation{Department of Materials Science and Engineering, University of Florida}

\date{\today}

\begin{abstract}

The inverse design of materials with specific desired properties, such as high-temperature superconductivity, represents a formidable challenge in materials science due to the vastness of chemical and structural space. We present a guided diffusion framework to accelerate the discovery of novel superconductors. A DiffCSP foundation model is pretrained on the Alexandria Database and fine-tuned on 7{,}183 superconductors with first-principles–derived labels.
Employing classifier-free guidance, we sample 200{,}000 structures, which lead to 34{,}027 unique candidates. 
A multistage screening process that combines machine learning and density functional theory (DFT) calculations to assess stability and electronic properties, identifies 773 candidates with DFT-calculated $T_\mathrm{c}>5$~K. Notably, our generative model demonstrates effective property-driven design.
Our computational findings were validated against experimental synthesis and characterization performed as part of this work, which highlighted challenges in sparsely charted chemistries.
This end-to-end workflow accelerates superconductor discovery while underscoring the challenge of predicting and synthesizing experimentally realizable materials.

\end{abstract}

\maketitle

\section{Introduction}
The discovery of novel materials with desired properties remains a fundamental challenge: traditional routes—direct simulation or experimental synthesis and characterization—are costly, slow, and yield few successes. While first principles methods and machine-learning (ML) models have improved property prediction from known crystal structures, the inverse problem—designing materials for target properties—remains daunting due to the vastness of chemical and structural space.

Generative models, which have achieved remarkable success in domains such as images, text, and video, are now gaining traction in materials science. Recent advances in structure-generating models, such as diffusion-based frameworks \cite{ho2020denoising} (e.g., DiffCSP \cite{jiao2023crystal}, MatterGen \cite{zeni2024mattergengenerativemodelinorganic}), flow-based \cite{lipman2023flow} models (e.g., FlowMM \cite{miller2024flowmm}), autoregressive transformers \cite{vaswani2017attention} (e.g. CrystalLLM \cite{crysgen}, Matra-Genoa \cite{wygen}), and frameworks based on stochastic interpolants \cite{albergo2023stochasticinterpolantsunifyingframework} that unify diffusion and flow (e.g., OMatG \cite{hollmer2025open}), highlight the potential of generative approaches for material discovery.

Superconductors, with their zero electrical resistance, hold transformative potential for technologies ranging from energy transmission and storage to high-field applications in medical imaging, particle accelerators, and materials processing~\cite{ROADMAP, Larbalestier2001, 10.1063/1.1955478, doi:10.1142/S1793626812300010}.
These wide-ranging applications make the discovery of new superconductors, particularly those with higher critical temperatures ($T_\mathrm{c}$), a high-priority target for inverse materials design. In the dominant class of electron-phonon superconductors, $T_\mathrm{c}$ is governed by the coupling between electrons and lattice vibrations \cite{PhysRev.108.1175, osti_7354388}. Traditionally density functional theory (DFT) \cite{10.1063/5.0005082} coupled to Eliashberg theory \cite{Ponc2016} has been used to accurately predict the $T_\mathrm{c}$ of superconductors, albeit at too high a computational cost to be suitable for high-throughput screening. Machine‑learning implementations of Eliashberg theory \cite{Xie2019, Xie2022} and of electron–phonon coupling \cite{Gibson2025, gibson2025developingcompleteaiacceleratedworkflow} are potential routes to accelerate the prediction of superconducting properties and screening efforts, enabling high‑throughput screening.

\begin{figure*}[!t]
    \includegraphics[width=1.0\linewidth]{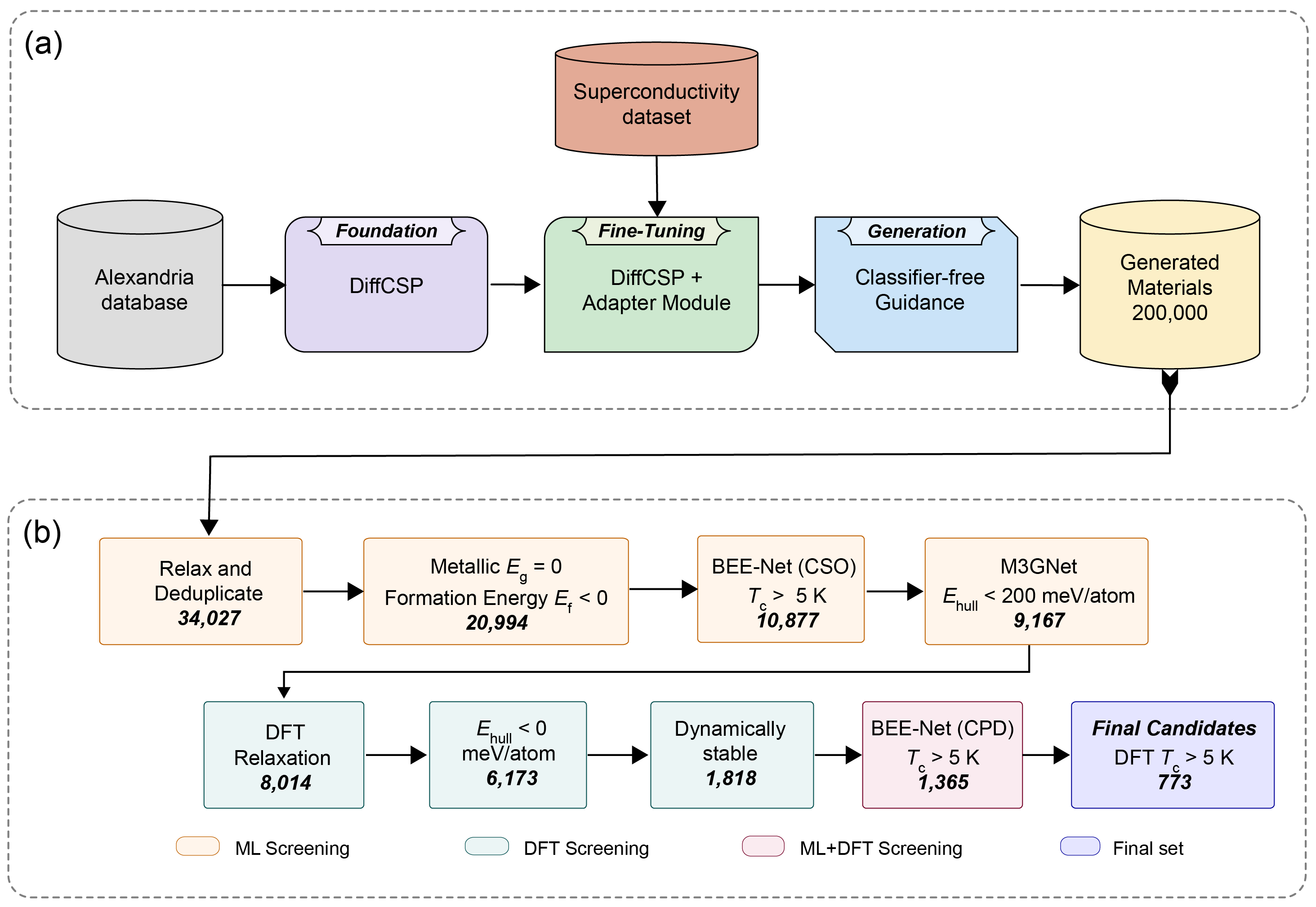}\\
    \caption{Workflow combining (a) guided diffusion and (b) multi-stage filtering to predict candidate superconductors.
    (a) Overview of the guided diffusion pipeline for superconductor discovery, starting from the Alexandria database of 2{,}086{,}767 crystal structures used here to train a DiffCSP‑based foundation model capable of generating plausible crystals. An adapter module is implemented in the DiffCSP denoiser to fine-tune the foundation model using a dataset of 7{,}183 superconductors \cite{Cerqueira2023, gibson2025developingcompleteaiacceleratedworkflow}.
    Using classifier free guidance, we use this AI framework to generate 200{,}000 crystal structures.
    (b) Overview of the multi-stage filtering process of the generated structures. We first pass the generated structures through initial ML relaxation and deduplication, followed by search for metallic, thermodynamically stable, ML predicted $T_\mathrm{c} > 5$ K structures and $E_\mathrm{hull} < 200$ meV/atom. We perform dynamic stability verification on the remaining candidates, and calculated the stable structures' $T_\mathrm{c}$ with BEE-NET \cite{gibson2025developingcompleteaiacceleratedworkflow}
    using the PhDOS as embedding, selecting the $T_\mathrm{c} > 5$ K. The electron-phonon spectral function of the final candidates is calculated using DFT,  which is then used to obtain the final (DFT) $T_\mathrm{c}$ of the candidate structures.}
    \label{fig:workflow}
\end{figure*}

Generative AI methods have the capability to directly propose crystal structures biased toward superconductivity by conditioning on target properties, potentially increasing the success rate relative to screening‑only workflows. Generative-AI efforts at superconductor discovery include Wines \textit{et~al.}~\cite{Wines_2023}, who combined a crystal diffusion variational autoencoder (CDVAE \cite{xie2022crystal}) with ALIGNN \cite{choudhary2021atomistic} as the property predictor. That study was a valuable proof of concept, but ALIGNN was trained on a small dataset (1{,}058 structures), limiting generalization, and the generated structures were reported in the low-symmetry space group $P\bar{1}$, which may bias diversity and physical realizability. Dordevic \textit{et~al.}~\cite{Yuan2024} introduced SuperDiff, a diffusion-based model that uses ILVR \cite{9711284} to condition on reference compounds; while fast and the first to generate new superconductor families via conditioning, it operates at the composition level without explicit crystal structures, limiting its ability to capture atomic arrangements. Other notable advances in AI‑driven materials-discovery include Pogue \textit{et al.}\ \cite{Pogue2023}, Wilfong \textit{et al.}\ \cite{Wilfong2025}, Hutcheon \textit{et al.}\ \cite{Hutcheon2020}, Chen \textit{et al.}\ \cite{mdl2021}, and Griesemer \textit{et al.}\ \cite{Griesemer2025}, with additional context reviewed in Ref.~\cite{Madika2025}.

The field still lacks a scalable generative framework for superconductors that solves the inverse design problem under data scarcity conditions. Here, we build on recent advances in generative AI and present an enhanced DiffCSP framework that uses guided diffusion to generate crystal structures conditioned on target $T_\mathrm{c}$ values. To address the limited size of labeled superconductivity datasets, we decouple structural priors from property conditioning: we first pretrain DiffCSP on more than two million crystal structures from the Alexandria Database~\cite{alex1, alex2}, teaching the model to generate plausible crystals independent of any target property; we then fine-tune our model for $T_\mathrm{c}$ conditioning on a smaller, high-quality set of 7{,}217 conventional superconductors with calculated $T_\mathrm{c}$ \cite{Cerqueira2023} (Fig.~\ref{fig:workflow}a). This two-stage strategy leverages a massive corpus to learn structural validity while requiring only a specialized dataset to capture superconductivity-relevant correlations, thereby steering generation toward candidates that are both plausible and likely to be superconducting.

We apply this end-to-end workflow at scale by generating 200{,}000 candidate crystal structures (Fig.~\ref{fig:workflow}a) and filtering them through a multistage screening process that combines machine-learning models \cite{doi:10.1021/acs.chemmater.9b01294, Chen2022, gibson2025developingcompleteaiacceleratedworkflow} with DFT \cite{10.1063/5.0005082}. The pipeline enforces metallicity ($E_f<0$), thermodynamic stability ($E_\mathrm{hull}<200$~meV/atom computed via ML and DFT, where $E_\mathrm{hull}$ is the energy above the hull), and dynamical stability (phonons), and prioritizes superconducting propensity. This process yields 773 novel candidates with DFT-calculated $T_\mathrm{c}>5$~K and $E_\mathrm{hull}<200$~meV/atom (Fig.~\ref{fig:workflow}b). Finally, we report synthesis and characterization for a subset to validate predictions and discuss practical criteria for designating genuine discoveries.


\section{Results}
We first validate our computational approach by demonstrating that guidance allows for the generation of materials with desired superconducting critical temperatures. We then characterize the full set of generated structures, followed by experimental synthesis of selected candidates for further study.

\subsection{Validation of the Guided Diffusion model}
We validated the performance of the guided diffusion model by assessing its ability to generate structures conditioned on a target $T_\mathrm{c}$. The guidance mechanism, based on classifier-free guidance \cite{ho2021classifierfree}, is designed to balance this drive toward a specific $T_\mathrm{c}$ with the principles of structural viability by the foundation model (Sec.~\ref{sec:foundation_model} and Sec.~\ref{sec:fine_tuning}). This balance is tuned by a guidance weight $w$ (detailed in Sec.~\ref{sec:cfg}). To validate our guided diffusion model we analyze the distributions of $T_\mathrm{c}$ values for sets of 1,000 generated structures, estimating $T_\mathrm{c}$ using crystal structure only (CSO) input to BEE-NET \cite{gibson2025developingcompleteaiacceleratedworkflow} 
We did not perform any structural stability test on the generated candidates for this step. This approach provides near-immediate feedback without the computational cost of ab initio calculations, which are reserved for the more comprehensive analysis workflow described in Sec.~\ref{sec:structure_analysis_methods}. As to the efficacy of DiffCSP to generate stable and novel structures, we refer the reader to Ref.~\onlinecite{jiao2023crystal}.

\begin{figure*}[t!]
    \centering
    \includegraphics[width=1.0\linewidth]{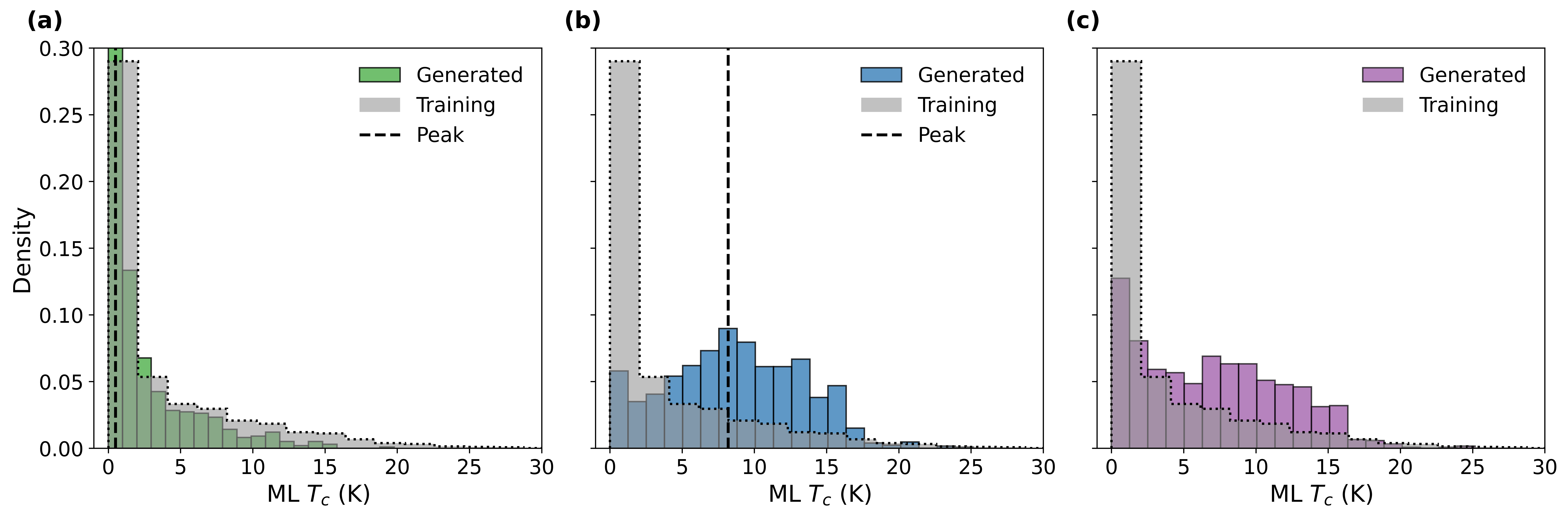}
    \caption{Effect of classifier-free guidance on $T_\mathrm{c}$ distribution of 1{,}000 generated structures per setting, with $T_\mathrm{c}$ predicted by BEE-NET with the crystal structure only (CSO) input. (a) Generated structures with guidance disabled ($w = -1$) closely follow the fine-tuning dataset distribution. (b) With guidance enabled ($w = 2$) and target $T_\mathrm{c} = 10$ K, the distribution shifts toward the desired value, demonstrating the model's controllability. (c) With $w = 2$ and target $T_\mathrm{c} = 110$ K - well outside the training distribution - the sampled distribution remains concentrated at low $T_\mathrm{c}$, indicating adherence to the learned plausible structural prior rather than oversteering to satisfy the property target.
}
    \label{fig:guidance_check}
\end{figure*}

Our analysis illustrated in Fig.~\ref{fig:guidance_check} first confirms that with guidance disabled ($w=-1$), the $T_\mathrm{c}$ distribution of the generated samples closely follows that of the fine-tuning dataset, establishing a baseline (Fig.~\ref{fig:guidance_check}a). In contrast, enabling classifier-free guidance (e.g., $w=2$) to target a specific $T_\mathrm{c}$ of 10~K results in a clear and effective shift of the distribution towards the desired value, demonstrating property-driven control (Fig.~\ref{fig:guidance_check}b).
We further probed the model behavior with an out-of-distribution target of $T_\mathrm{c} = 110$~K, a regime absent from the fine-tuning data. Even with strong guidance, the model consistently failed to produce high-$T_\mathrm{c}$ structures. Instead, as shown in  Fig.~\ref{fig:guidance_check}c ($w=2$), the model preferentially generates low-$T_\mathrm{c}$ structures. This behavior demonstrates that while guidance is effective for targeted design within the training domain, the foundation model acts as a crucial prior for stability, constraining the generation to physically plausible structures for extreme property extrapolation.

\subsection{Prediction of Superconducting Candidates}

We applied the full generative–screening pipeline to 200{,}000 initial structures. After fast MEGNet \cite{doi:10.1021/acs.chemmater.9b01294} relaxation and deduplication, 34{,}027 unique and novel generated structures remained. These were subjected to the multistage structural-analysis workflow detailed in Section~\ref{sec:structure_analysis_methods}.

\begin{figure*}[t]
    \centering
    \includegraphics[width=0.8\linewidth]{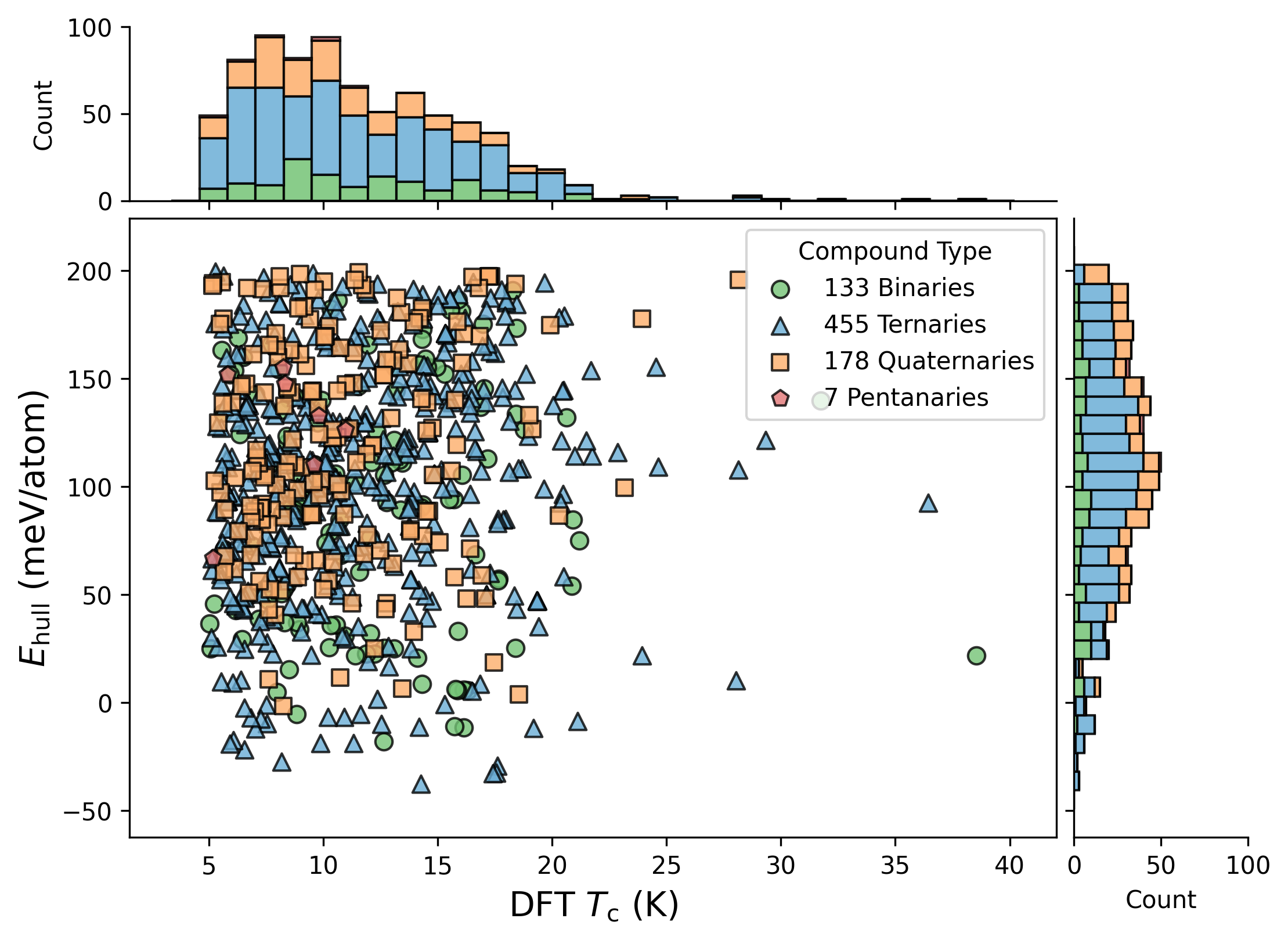}
    \caption{Distribution of the 773 predicted superconducting candidates. The main panel shows energy above the convex hull ($E_\mathrm{hull}$) versus DFT‑calculated $T_\mathrm{c}$; marker color/shape denotes compound type as indicated in inset legend. Marginal histograms along the top (for $T_\mathrm{c}$) and right (for $E_\mathrm{hull}$) display stacked counts by compound type, using 30 equal‑width bins aligned to the main panel axis limits.}
    \label{fig:candidates}
\end{figure*}

Of the 34{,}027 identified structures, 20{,}994 were classified as metals with negative formation energy ($E_f<0$). From this set, 10{,}877 were predicted to have $T_\mathrm{c}>5$~K by BEE-Net using crystal structure only (CSO) inputs. Thermodynamic pre-screening with M3GNet retained 9{,}167 with energy above the convex hull $E_\mathrm{hull}<200$~meV/atom. After DFT relaxation, 8{,}014 structures converged; 6{,}173 of these satisfied a DFT-calculated $E_\mathrm{hull}<200$~meV/atom \cite{Sun2016}. Phonon calculations identified 1{,}818 as dynamically stable. Incorporating coarse phonon-density (CPD) embeddings into BEE-Net yielded 1{,}365 with predicted $T_\mathrm{c}>5$~K, from which 773 unique candidates exhibit DFT-calculated $T_\mathrm{c}>5$~K. The relationship between predicted $T_\mathrm{c}$ and $E_\mathrm{hull}$ for these candidates is shown in Figure~\ref{fig:candidates}, spanning $T_\mathrm{c}$ values up to $\sim$35~K and highlighting materials that combine promising superconducting properties with thermodynamic stability conducive to synthesis.

A compositional analysis of these 773 candidates revealed a strong trend towards multi-component compounds with 133 binaries (17\%), 455 ternaries (59\%), 178 quaternaries (23\%), and 7 pentanaries (1\%). We attribute this trend to the combined effect of our model's generative capabilities and the current state of superconductor research. The landscape of binary superconductors has been extensively investigated. In essence, for the workflow to identify new and viable superconductors, it is more probable that these will emerge from these more complex systems where a greater potential for undiscovered materials exists, rather than from the more saturated binary space.

Our generative AI approach proved to be highly effective, as out of 34,027 unique structures 773 are predicted to be superconductors with $T_\mathrm{c}>5K$. For comparison, an element‑substitution baseline approach using the same BEE‑Net/DFT workflow identified 204 superconductors from 1.22 million unique candidates (0.017\%), whereas the here-proposed generative pipeline yields 773 of 34{,}027 (2.3\%), a $\sim$135$\times$ improvement in hit rate \cite{gibson2025developingcompleteaiacceleratedworkflow}. Within generative approaches, Wines et al.\ generated 3{,}000 candidates and obtained 25 materials with DFT‑calculated $T_\mathrm{c}>5$~K (0.83\%) \cite{Wines_2023}; the present pipeline achieves a higher per‑structure hit rate (2.3\% vs.\ 0.83\%, $\sim$2.8$\times$) and a larger absolute yield (773 vs.\ 25, $\sim$31$\times$).

\subsection{Experimental Exploration}
Additional criteria are needed to down-select the list of 773 compounds to a manageable number of candidates for synthesis.
We decided to concentrate our initial efforts on ternary systems.
Binary systems were deliberately excluded due to extensive prior explorations~\cite{gibson2025developingcompleteaiacceleratedworkflow}, significantly diminishing the likelihood of discovering new binary superconductors.
Quaternary and higher systems were also temporarily set aside because many of the proposed candidates had compositions that appeared quite similar to known high-entropy and multi-principal component alloys.
Hence, restricting our target materials to ternaries reduced our list of candidates to 455 compounds.
Further, we ruled out materials candidates containing the radioactive element technetium, leaving 315 compounds.

With the list of candidates still so large, time-consuming and complicated synthesis methods such as flux growth or solid-state reaction with multiple grinding and pelletizing stages remained impractical.
We therefore settled on arc melting as a rapid and straightforward means to synthesize several candidates.
Not all combinations of elements are suitable for arc melting.
Generally, one must ensure that vapor pressures of the constituent elements do not become too high at the melting point of the element with the highest melting temperature.
Based on vapor pressure data~\cite{HonigKramer1969}, we developed a screening function that allowed us to further downselect to materials that should be easily arc meltable without carefully accounting for vaporization losses,
resulting in 51 compounds. 

We noticed that a number of these materials appeared to be substituted variations of the A15 structure.
These were systematically identified by attempting every permutation of reducing each ternary to a binary compound and then comparing the resulting structures against the Cr$_3$Si (A15) prototype using the \texttt{StructureMatcher} module within \texttt{pymatgen}.
We decided against further investigation of materials that were substituted A15 variants due to the extensive prior work on these materials over several decades.
As one example, our model predicted a substituted A15 variant Nb$_6$GaSi with predicted $T_\mathrm{c}$ of 36 K (and 90 meV above the hull).
In 1982, Nb$_3$(Ga,Si) was investigated in detail, finding $T_\mathrm{c}$ values below the $\sim \SI{21}{K}$ reported for pure Nb$_3$Ga~\cite{Herold1982}.
Leaving out substituted A15 variants resulted in 46 compounds.

To maximize the likelihood of synthesizability, one would of course like to select only materials with energies on or below the convex hull.
Applying this additional criterion further culls the list to only two compounds: MoNbTa$_2$ and MoNb$_2$Ta.
We attempted to synthesize both of these structures; the results are detailed below.
Given the small number of materials meeting this very strict criterion, we also attempted to synthesize several materials with energies predicted to be close to, but above, the convex hull.

To prioritize synthesis attempts among these remaining 44 materials, we compared all structures pairwise again using the \texttt{StructureMatcher} module of \texttt{pymatgen} \cite{pymatgen}.
Structures were then clustered into groups using a graph-based approach via the NetworkX library \cite{networkx}.
Through this method we identified several structures that appeared multiple times with different elements.
In order to diversify the materials chosen for synthesis, we prioritized selecting one or two candidates from each structure family.

Ultimately, this comprehensive screening and clustering approach yielded a final selection of 18 structurally diverse and energetically promising candidates designated for synthesis and characterization.
In order to conserve limited experimental resources and time, our experimental workflow also used a multi-stage screening approach.
Each sample was initially subjected to a rapid ``dip-stick'' AC magnetic susceptibility measurement that determines only if the sample is superconducting above \SI{4.2}{K} (and a rough shielding fraction) without providing any specific information on the exact $T_\mathrm{c}$ value.
These measurements take only about 10 minutes.
Only samples that showed positive evidence for superconductivity in the dip stick measurements were subjected to additional and more time-consuming temperature-dependent AC magnetic susceptibility measurements to pinpoint the value of $T_\mathrm{c}$ and X-ray diffraction measurements to determine if the predicted structure formed.

\begin{table*}[t]
\centering
\scriptsize
\setlength{\tabcolsep}{1pt}
\begin{ruledtabular}
\begin{tabular}{llllllll}
\# & Formula & \makecell{Predicted \\ $T_\mathrm{c}$ (K) \\ ($\mu^*=0.10$)} &
\makecell{Predicted \\ $T_\mathrm{c}$ (K) \\ ($\mu^*=0.21$)} &
\makecell{Measured \\ $T_\mathrm{c}$ (K) Onset} &
\makecell{$E_\mathrm{hull}$ \\ (meV)} &
X-Ray & Relevant known superconductors\\
\hline
1  & MoTaTi$_2$       & 16.5 K & 9.47 K & --   & 62  & --                 & -- \\
2  & MoTa$_2$Ti       & 13.9 K & 7.49 K & --   & 70  & --                 & -- \\
3  & HfTa$_4$Zr & 13.7 K & 9.57 K & 7.3 K & 115 & bcc solid solution & \makecell[l]{bcc Ta--Zr ($T_\mathrm{c} \lesssim \SI{7.7}{K}$)~\cite{Linker1983} \\ and bcc Hf--Ta ($T_\mathrm{c} \lesssim \SI{6.9}{K}$)~\cite{Gey1972}} \\
4  & HfMoTa$_2$       & 12.7 K & 7.36 K & --   & 115 & --                 & -- \\
5  & MoTaZr$_2$       & 11.3 K & 6.74 K & 5.9 K & 49  & two bcc solid solutions & bcc Ta--Zr ($T_\mathrm{c} \lesssim \SI{7.7}{K}$)~\cite{Linker1983} \\
6  & Hf$_2$MoTi       & 10.5 K & 5.49 K & --   & 105 & --                 & -- \\
7  & HfMoTi           & 10.5 K & 5.15 K & --   & 82  & --                 & -- \\
8  & Nb$_2$TaZr & 19.6 K & 13.69 K & 9.7 K & 99  & bcc solid solution & \makecell[l]{bcc Nb--Ta ($T_\mathrm{c} \sim 4$--\SI{9}{K}) \\ and bcc Nb--Zr ($T_\mathrm{c} \sim 8$--\SI{11}{K})~\cite{Hulm1961}} \\
9  & NbTa$_2$Zr & 15.2 K & 9.34 K & 8.8 K & 108 & bcc solid solution & \makecell[l]{bcc Nb--Ta ($T_\mathrm{c} \sim 4$--\SI{9}{K}) \\ and bcc Nb--Zr ($T_\mathrm{c} \sim 8$--\SI{11}{K})~\cite{Hulm1961}} \\
10 & Hf$_2$MoNb       & 10.2 K & 5.12 K & --   & 54  & --                 & -- \\
11 & MoNbZr$_2$       & 11.4 K & 5.59 K & 5.7 K & 101 & bcc solid solution & bcc Nb--Zr ($T_\mathrm{c} \sim 8$--\SI{11}{K})~\cite{Hulm1961} \\
12 & Mo$_3$NbTi$_2$   & 10.8 K & 4.29 K & --   & 30  & --                 & -- \\
13 & Hf$_2$Nb$_3$Ru   & 9.7 K  & 4.00 K & 6.8 K & 41  & multiphase         & bcc Hf--Nb ($T_\mathrm{c} \sim 5$--\SI{10}{K})~\cite{Siemens1969_RI7258} \\
14 & Mo$_3$NbRu$_2$ & 6.9 K & 1.77 K & 5.1 K & 135 & Mo$_2$Ru + bcc solid solution & \makecell[l]{bcc Mo--Nb ($T_\mathrm{c} \sim 3$--\SI{9}{K})~\cite{Hulm1961}, \\ Mo$_2$Ru not characterized} \\
15 & MoNbRu$_2$ & 9.3 K & 6.05 K & 4.8 K & 140 & Mo$_2$Ru + unknown & \makecell[l]{bcc Mo--Nb ($T_\mathrm{c} \sim 3$--\SI{9}{K})~\cite{Hulm1961}, \\ Mo$_2$Ru not characterized} \\
16 & Re$_4$TaZr       & 18.7 K & 12.86 K & 5.25 K & 108 & Re$_2$Zr + hcp   & Re$_2$Zr ($T_\mathrm{c}=\SI{6.4}{K}$)~\cite{Giorgi1970} \\
17 & MoNbTa$_2$     & 7.0 K  & 2.30 K & --   & -12 & bcc solid solution & -- \\
18 & MoNb$_2$Ta     & 6.6 K  & 1.73 K & --   & -2  & bcc solid solution & -- \\
\end{tabular}
\end{ruledtabular}
\caption{Selected from the 773 candidates identified by our generative–screening workflow, this table lists experimentally synthesized and characterized materials with their respective predicted and measured $T_\mathrm{c}$ values. Predicted values are obtained from DFT electron–phonon calculations via the Allen–Dynes formula with Coulomb pseudopotentials $\mu^*=0.10$ and $\mu^*=0.21$ \cite{gibson2025developingcompleteaiacceleratedworkflow}. Measured $T_\mathrm{c}$ values reported as "–" correspond to samples that did not show signs of superconductivity above \SI{4.2}{K}.}
\label{tab:synthesizedList}
\end{table*}

\begin{figure*}
    \centering
    \includegraphics[width=1.0\textwidth]{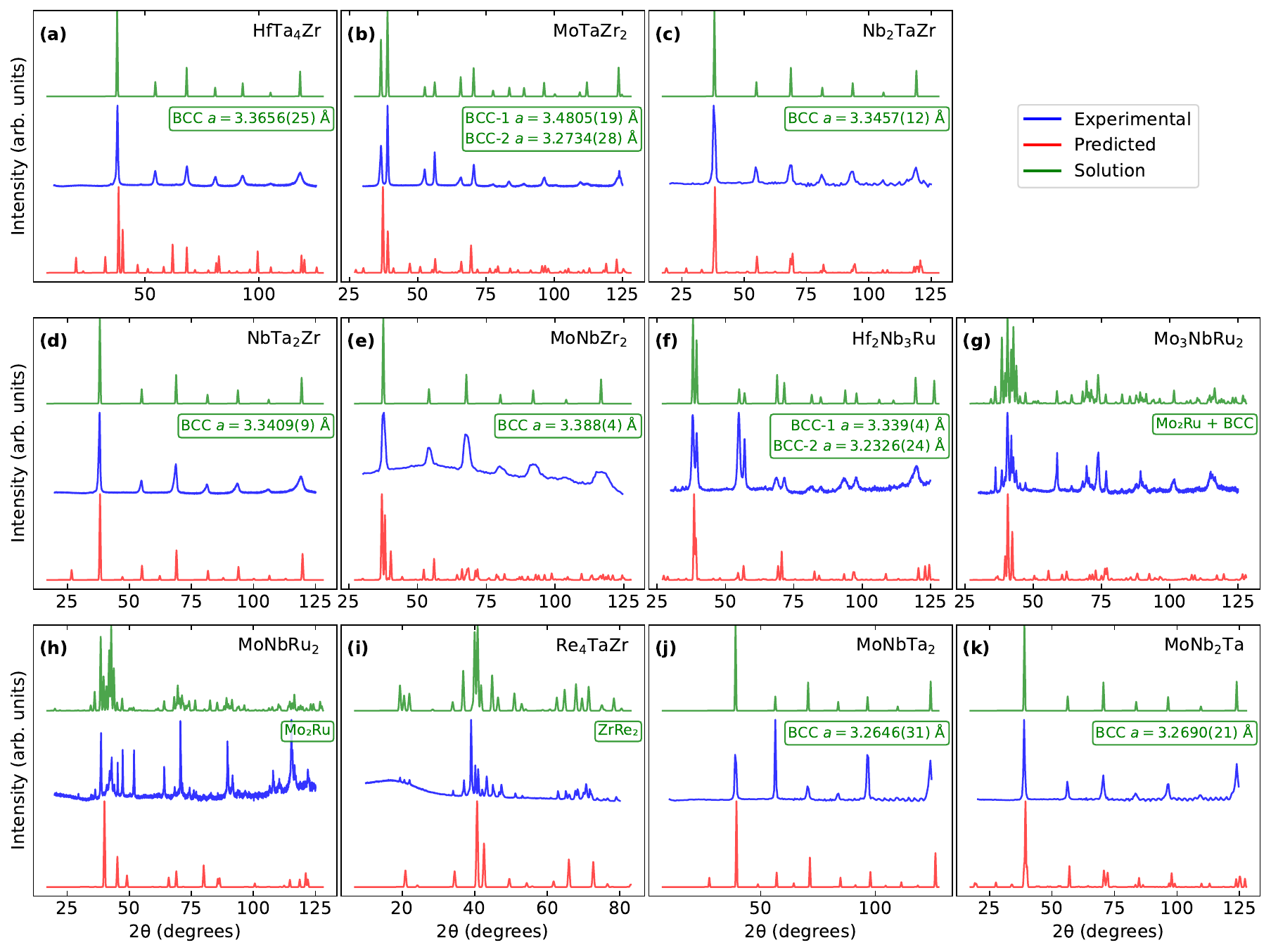}
    \caption{
    X-ray diffraction data for eleven samples, including samples that either exhibited evidence for superconductivity above \SI{4.2}{K} (panels a - i) or were predicted to be on the hull (samples j and k).
    The blue curves are experimental data (after background subtraction), red curves show a simulated pattern for the structure predicted by our model, and the green curves show a simulated pattern for our solution to the structure (or mixture of structures).
    Although there are some cases where the data (blue) shows gross similarities to the predicted (red) pattern, in every case, the data is much better fit by a simple disordered BCC structure, a known binary phase, or a combination thereof.
    In several cases (e.g. panels a, c, d, j, and k) the absence of predicted peaks in the lowest angle regime strongly confirms that the data is better descibed by simple disordered BCC alloys rather than the complex ternary ordered compounds predicted by the model.
    }
    \label{fig:xrd}
\end{figure*}

Table~\ref{tab:synthesizedList} presents a summary of the experimental results.
A dash (``-'') in the ``Measured $T_\mathrm{c}$ (K) Onset'' column indicates that no evidence for superconductivity was detected above \SI{4.2}{K}.
Of the 18 materials, 9 showed evidence for superconductivity above \SI{4.2}{K}, although the $T_\mathrm{c}$ values were consistently lower than predicted. 
While these initial results appear highly promising, X-ray diffraction measurements revealed that, in general, the predicted structures did not form, and instead we synthesized mostly disordered BCC solid solutions or multi-phase mixtures of known binary phases and simple BCC solid solutions.
Figure~\ref{fig:xrd} presents an overview of the X-ray diffraction data.

\section{Discussion}

In this work, we developed and deployed an end-to-end workflow that combines a guided diffusion generative model with high-throughput computational screening to accelerate the inverse design of novel superconductors.
As quantified in the Results, our pipeline achieves a 2.3\% per‑structure hit rate (773/34{,}027), improving on an element‑substitution baseline (0.017\%, 204/1.22\,million \cite{gibson2025developingcompleteaiacceleratedworkflow}) by $\sim$135$\times$; relative to a recent generative search (25/3{,}000 \cite{Wines_2023}), it also delivers a $\sim$31$\times$ larger absolute yield (773 vs.\ 25).

The distribution of stability, i.e.\ $E_\textrm{hull}$, as shown in Fig.\ref{fig:candidates}, can likely be attributed to the pre-training and fine-tuning datasets, both consist mostly of theoretical materials that are not all on the convex hull.
It is likely that training on experimentally synthesized materials, and/or only materials with $E_\textrm{hull}$ near zero, would guide the model to generate candidates that are, on average, closer to the hull.

As noted in the Results section, the final candidate set skews toward ternary compositions, reflecting the extensive prior exploration of binaries. Many of these ternaries involve relatively costly elements, suggesting historical under‑sampling of such chemistries in experiments. Future iterations should encode cost and handling constraints (e.g., excluding toxic or radioactive elements) as generative priors and screening filters, steering discovery toward candidates that are both scientifically promising and experimentally tractable. At the same time, pushing into underexplored chemistries raises a distinct caveat we discuss below: data sparsity can make phases appear stable relative to an incomplete convex hull.

A critical limitation arises in underexplored chemistries that may partly explain our experimental outcomes. The model's prediction of stable phases in these regions could be an artifact of data scarcity, not an indication of true physical plausibility. For a predicted structure to be deemed stable, its energy must be low relative to a convex hull constructed from known, competing phases. In chemical spaces where data is sparse -- for instance, among ternary systems whose constituent elements do not readily form any known binary compounds -- the reference convex hull is likely incomplete. Consequently, a generated structure may appear to have a low energy above the hull ($E_\mathrm{hull}$) simply because there are no known, more stable structures to compete with it in the database. Our generative model, therefore, may be expertly identifying gaps in existing materials data rather than discovering genuinely synthesizable, stable phases. Indeed, close to two thirds of the binary phase diagrams that are relevant to the materials predicted by the model contain no known ordered binary compounds, often because the elements form complete solid solutions.

The factors above highlight a fundamental challenge for AI-driven materials discovery: distinguishing true, synthesizable novelty from artifacts of an incomplete reference dataset.
Therefore, while our work validates the immense potential of generative AI to accelerate materials discovery, it also underscores the critical need for future developments to focus on improving predictions of synthesizability and integrating experimental feedback into a true active learning loop to bridge the gap between computational prediction and laboratory realization.

\begin{figure}
    \centering
    \includegraphics[width=\linewidth]{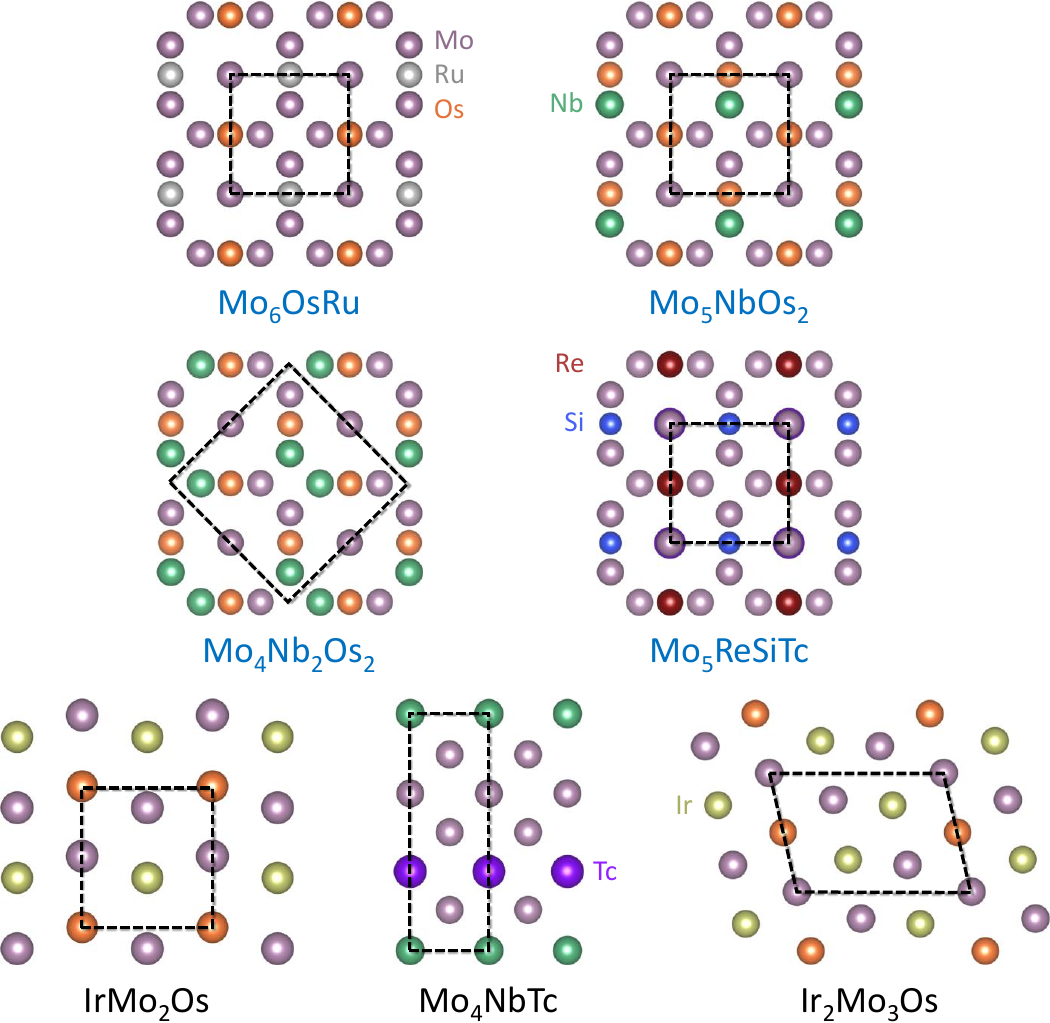}
    \caption{
    Seven novel structures generated by the model that do not match any known prototype in the AFLOW prototype encyclopedia~\cite{Eckert2024}.
    Compounds written in blue can be considered partially substituted derivatives of the A15 structure.
    }
    \label{fig:prototypes}
\end{figure}

Among the initial dataset of 773 crystal structures, there were 28 unique stable (on the hull) structures. These structures were compared to the AFLOW prototype encyclopedia using AFLOW-XtalFinder~\cite{Hicks2021}. Of the 28 structures, 8 matched known prototypes in the AFLOW database, 5 represented novel arrangements of known symmetry frameworks (possessing isopointal prototypes), and 15 were completely novel with no corresponding prototypes or isopointal matches in the encyclopedia. The 15 completely novel stable structures were subjected to clustering analysis using AFLOW's structure comparison algorithms. This clustering identified 7 distinct structural families among the 15 structures. The largest family contained 9 structures, all adopting a space group 200 (cubic $\mathrm{Pm}\overline{3}\mathrm{m}$) structure that is a substituted variant of the Cr$_3$Si A15 structure type.
The remaining 6 families were represented by single structures.
The structures are shown in Fig.~\ref{fig:prototypes}, where the four compounds indicated in blue can be considered as substituted derivatives of the A15 structure.

To avoid overcounting elemental re‑decorations and potential training‑set reproduction—issues raised in subsequent commentary on generative crystal design \cite{Juelsholt2025}—we adopt a conservative novelty criterion: unique after deduplication, no prototype or isopointal match in the AFLOW prototype encyclopedia (AFLOW‑XtalFinder), and passing stability screens (DFT‑relaxed, $E_\mathrm{hull}<200$~meV/atom, dynamically stable where computed).
These prototypes provide a potential starting point for systematic element substitution to identify stable ordered compounds \cite{Griesemer2021} that are ternary derivatives of the A15 structure. However, the potential for disorder must carefully considered to avoid predicting compounds that instead form as solid solutions as in the case of Nb$_3$(Ge, Si)~\cite{Herold1982} that was discussed above.

Future directions could involves improvements to the generative model itself.
Replacing fixed property embeddings with learnable ones and evaluating alternative guidance schemes \cite{karras2024guiding, tang2025diffusionmodelsclassifierfreeguidance} may improve generalization and coverage relative to classifier-free guidance.
To better reflect the discrete nature of the composition, guided discrete diffusion for element identities \cite{nisonoff2025unlocking} may be preferable over denoising in a continuous latent space.
It is also natural to integrate our guidance framework with alternative generative backbones—including Flow Matching \cite{miller2024flowmm} and stochastic interpolants \cite{hollmer2025open}—to assess their efficacy in discovering novel, synthesizable superconductors.

Bridging prediction and synthesizability points to several clear next steps. To address sparse reference data in undercharted chemical spaces, on-the-fly DFT refinement of local convex hulls around promising candidates could be employed. An experiment-in-the-loop active-learning workflow—incorporating non-formations as hard negatives, retraining the model iteratively, and expanding training data with experimentally verified compounds—could sharpen decision boundaries. Multi-objective optimization that balances high $T_\mathrm{c}$ with synthesizability and practical constraints (cost, handling, toxicity) could further focus searches. Finally, disorder-aware screening that compares predicted ordered compounds against competing solid solutions could improve experimental realizability and hit rates.


\section{Methods}
\label{sec:methods}
Our methodology for discovering novel superconductors integrates three primary components: (1) a generative AI framework to propose novel crystal structures conditioned on superconducting properties; (2) a high-throughput computational screening workflow driven by machine learning models and density functional theory (DFT) to assess the stability and promise of the generated candidates; and (3) experimental synthesis and characterization efforts to validate the most promising predictions. The subsections below detail the computational and experimental procedures.

\subsection{Computational Methods}
The computational pipeline (Fig.~\ref{fig:workflow}) first generates a large ensemble of candidates and then systematically filters them to identify the most promising ones. The generative stage comprises three steps: pretraining a foundation model, fine-tuning it for superconductivity, and sampling with classifier-free guidance to generate 200{,}000 candidate structures (Fig.~\ref{fig:workflow}a). These candidates are then subjected to a rigorous, multi-stage filtering workflow (Fig.~\ref{fig:workflow}(b)) that uses a combination of machine learning interatomic potentials \cite{doi:10.1021/acs.chemmater.9b01294, Chen2022}, BEE-NET property predictor \cite{gibson2025developingcompleteaiacceleratedworkflow}, and DFT calculations \cite{10.1063/5.0005082, vasp1, vasp2} to assess metallicity, thermodynamic and dynamical stability, and to estimate $T_\mathrm{c}$. This process distills the large initial set to a refined list of novel, stable candidates suitable for experimental consideration.

\subsubsection{Data}
We used two datasets: (1) an unlabeled corpus from the Alexandria Materials Database \cite{alex1, alex2} for pretraining, and (2) a labeled superconductors set for fine-tuning. The Alexandria database contains over five million crystal structures; for pretraining we selected all cells with $\leq$20 atoms per unit cell, yielding 1{,}857{,}222 structures for training and 229{,}545 for validation. This large corpus supplies a structural prior independent of superconducting properties.

For fine-tuning, we used the DS-A dataset of Cerqueira et al.\ \cite{Cerqueira2023}, comprising 7{,}217 dynamically stable metallic compounds with first-principles electron–phonon results (relaxed structures, density of states at the Fermi level, logarithmic average phonon frequency, and electron–phonon coupling). We used the Allen–Dynes formula \cite{AllenDynes1975} to estimate $T_\mathrm{c}$ of these materials, and retaining 6{,}326 superconducting materials, reduction due to the numerical instability from the Allen-Dynes formula. We then augmented this set with 857 additional superconducting candidates from \cite{gibson2025developingcompleteaiacceleratedworkflow}, yielding 7{,}183 labeled samples for fine-tuning. The fine-tuning data were split into training/validation/test in an 8:1:1 ratio, and early stopping was used during training. We use only the $T_\mathrm{c}$ values for fine-tuning. 

\subsubsection{Foundation Model Training}
\label{sec:foundation_model}
The first stage in our methodology is the development of a robust foundation model for crystal structure generation.
For this, we trained a diffusion-based generative model with the Alexandria Materials Database.
The goal of this stage was to learn a broad structural prior --- i.e., to generate diverse, plausible crystal structures --- without conditioning on superconducting properties. Training followed the DiffCSP framework
(architecture in Supplementary information section S1).
In this unsupervised phase, the model learns to reverse the diffusion process applied to lattice vectors, fractional atomic coordinates, and element types.
By learning to effectively denoise these inputs, the model implicitly learns the complex manifold of plausible  crystal structures. This process is guided by the need to respect fundamental physical symmetries inherent in crystalline materials, such as rotational equivariance and periodic boundary conditions. The resulting pre-trained model is therefore well-versed in the general principles of crystal structure formation, without being biased towards any specific target property, providing a strong starting point for subsequent fine-tuning towards superconductor discovery. We kept the model hyperparameters consistent with those used in the original DiffCSP framework.

\subsubsection{Fine-tuning for Conditional Generation}
\label{sec:fine_tuning}
The second stage of training involves fine-tuning the foundation model towards generation of superconducting materials. We introduce an adapter module \cite{zeni2024mattergengenerativemodelinorganic} into the equivariant graph network (CSPNet) used in DiffCSP; similar ideas have already been implemented in computer vision for text-to-image generation \cite{latentdiffusion, t2icontrol, t2iadapter}. This module allows us to condition generation on the target property, in this case, scalar $T_\mathrm{c}$ values. The adapter module is applied after each message-passing layer, at layer $L$, node embeddings are updated as
\[
H^{'(L)}_j =  H^{(L)}_j + f^{(L)}_{\text{mixin}}(f^{(L)}_{\text{adapter}}(g)) \cdot I(\text{property is not null})
\]
where ${H}_j^{(L)}$ denotes the embedding of node $j$ at layer $L$, and $g$ is a sinusoidal embedding \cite{vaswani2017attention} of the scalar target $T_\mathrm{c}$. The adapter $f_{\text{adapter}}^{(L)}$ consists of a small stack of four fully connected layers. Its output is passed through a mixin layer \cite{harrison2001mixin} $f^{(L)}_{\text{mixin}}$ that gates the adapter signal, stabilizing training by gradually increasing the adapter's influence. The mixin layer initially scales the adapter output to zero and increases it progressively during fine-tuning. This mechanism prevents abrupt disruptions to the foundation model’s learned features. 

We retain the original DiffCSP denoising objective, so the pretrained structural prior is preserved while the adapters inject $T_\mathrm{c}$ information. This yields a property-conditioned generator without degrading the foundation model’s learned notions of structural plausibility.

\subsubsection{Classifier-free Guided Generation}
\label{sec:cfg}
To generate candidate superconducting materials, we implemented classifier-free diffusion guidance \cite{ho2021classifierfree}. This method enables conditional generation by interpolating between the unconditional foundation model and the fine-tuned, property-aware model. Specifically, we combine the denoising predictions from both models as follows:
\[
\tilde{\epsilon}_{\theta}(z_{\lambda}, c) = (1 + w)\epsilon_{\theta}(z_{\lambda}, c) - w\epsilon_{\theta}(z_{\lambda})
\]
Here, $\epsilon_{\theta}(z_{\lambda}, c)$ is the denoising output of the conditional model guided by the target property $c$, and $\epsilon_{\theta}(z_{\lambda})$ is the output of the unconditional model. The guidance weight $w$ controls the strength of the conditioning — higher values bias the generation more strongly toward the target property. We found $w = 2$ to be a stable and effective choice in practice.

Using this guided generation strategy, we sampled 200,000 crystal structures by conditioning on different $T_\mathrm{c}$ values. These generated candidates were then passed through our structure analysis workflow to identify stable, high-$T_\mathrm{c}$ superconductors.

\subsubsection{Structural Analysis Workflow}
\label{sec:structure_analysis_methods}
We follow a rigorous multi-step workflow to identify stable, high-$T_\mathrm{c}$ superconducting candidates from the set of generated structures. This pipeline ensures that final materials are metallic, thermodynamically and dynamically stable, and synthesizable. Before these generated structures enter the detailed screening pipeline, an initial filtering step is performed to ensure a focus on novel candidates. We relax all structures using the M3GNet machine learning interatomic potential \cite{Chen2022} and duplicate structures within the generated set itself are identified and reduced to unique instances. Additionally, we remove any generated structures that are identical to those present in our fine-tuning dataset. This pre-screening ensures that the subsequent computationally intensive analyses are concentrated on genuinely new potential superconductors.

First, we use MEGNet \cite{doi:10.1021/acs.chemmater.9b01294} to compute the bandgap ($E_\mathrm{g}$) and formation energy ($E_f$). We retain only metallic structures ($E_\mathrm{g} = 0$) with negative $E_f$. Next, we estimate the superconducting critical temperature ($T_\mathrm{c}$) of these filtered structures using BEE-NET \cite{gibson2025developingcompleteaiacceleratedworkflow}. We only keep materials with predicted $T_\mathrm{c} > 5$ K for further analysis.
To assess thermodynamic stability, we calculate the energy above the convex hull ($E_\mathrm{hull}$) using M3GNet and compare against the Materials Project database \cite{jain2013materials}. Structures with $E_\mathrm{hull} > 200$ meV/atom are discarded. We further refine the surviving structures by recalculating $E_\mathrm{hull}$ using DFT-relaxed geometries and reapply the $E_\mathrm{hull} > 200$ meV/atom filter.

At this point, we incorporate phonon calculations to improve both dynamic stability assessment and $T_\mathrm{c}$ accuracy. For each structure, we compute the phonon density of states (PhDOS), ones with imaginary phonons are discarded and the remaining structures use BEE-Net with coarse phonon density (CPD) embeddings to predict a more refined $T_\mathrm{c}$. In the final stage, we perform electron-phonon coupling calculations using Quantum ESPRESSO to obtain the electron-phonon spectral function $\alpha^2F(\omega)$ and compute the final DFT-based $T_\mathrm{c}$ values using the Allen-Dynes equation \cite{AllenDynes1975}. 

\subsection{Experimental Methods}
Samples were arc-melted using either an Edmund-B\"uhler MAM-1 arc melter, or a custom built arc melter.
Both utilize a water-cooled copper hearth.
Samples were melted 3-5 times, flipping them over between each melting.
Mass losses were typically below 1\%.

Initial screening for superconductivity above \SI{4.2}{K} was performed using a custom-built dipstick probe made for performing AC magnetic susceptibility measurements while the sample is gradually inserted directly into a liquid He dewar.
The coils were calibrated using a piece of Pb of known size, which allowed us to calculate the approximate superconducting volume fraction for subsequent measurements.
The dip stick probe is not equipped with a thermometer, so the measurement only determines if the sample is superconducting above \SI{4.2}{K}.

For samples that showed evidence of superconductivity above \SI{4.2}{K} via dip stick measurements, more detailed AC magnetic susceptibility measurements were made using a Quantum Design Physical Property Measurement System (PPMS), a Stanford Research Systems SR830 Lock-In Amplifier, and custom wound coils of the type described in Ref.~\cite{Hamlin_dissertation}.

X-ray diffraction measurements were carried out using a Panalytical X'pert machine at The Nanoscale Research Facility at the University of Florida. 
The measurements utilized Cu-K$_\alpha$ radiation.
The angle step for the measurement is 0.0167$^\circ$ and collection time for each run is 80 minutes.
The samples include both powder and polished bulk forms depending on whether a given sample was brittle enough to grind into powder.
The GSAS-II software package was used for analysis of the XRD data.

\section*{Data availability}
All data supporting the findings of this study are available at within the paper and its Supplementary Information as well as at \url{https://huggingface.co/datasets/paprakash/GuidedMatDiffusion_data}.

\section*{Code availability}
Code used for model training and guided sampling is available at \url{https://github.com/paprakash/GuidedMatDiffusion}.

\section*{Acknowledgements}
This work was supported by the U.S. National Science Foundation under Grants OAC-2311632 (PP, EF, SM, ET, ML, AR, RH) and DMREF-2118718 (BG, ZL, GG, JE, JSK, GS, PH, JH, RH). Computational resources were provided by the University of Florida HiPerGator high‑performance computing system and the UF AI and Complex Computational Research program.

\section*{Author contributions}
The project was conceived by PP, PH, SM and RH. PP led the technical execution of the project, developed the guided‑diffusion framework, adapter conditioning, implemented code with contributions by EF, trained the generative model, generated materials and curated the candidate database. JG implemented the screening ML and DFT tools. ZL, GG, JE, GS, and JH carried out experimental synthesis and characterization. PH and RH supervised the research and provided guidance on analysis and interpretation. PP and JH drafted the manuscript; all authors discussed results and contributed to revisions.

\bibliography{references}

\end{document}